\documentclass[twocolappendix,appendixfloats,]{emulateapj}

\usepackage[colorlinks = true, linkcolor = blue, urlcolor  = blue, citecolor = blue, anchorcolor = blue]{hyperref}
\usepackage{hyperref}
\usepackage{apjfonts}
\usepackage{rotating}
\usepackage{amsmath}
\usepackage{color}
\usepackage{graphicx}
\usepackage[dvipsnames]{xcolor}
\usepackage{CJK}

\newcommand{\te}{t_{\rm E}}
\newcommand{\thetae}{\theta_{\rm E}}

\newcommand{\pie}{\pi_{\rm E}}

\newcommand{\dl}{D_{\rm L}}

\definecolor{brown}{rgb}{0.59, 0.29, 0.0}
\definecolor{darkgreen}{rgb}{0.0, 0.42, 0.24}
\definecolor{darkblue}{rgb}{0.01, 0.31, 0.59}
\definecolor{darkpurple}{rgb}{1.25, 0.38, 2.05}

\shorttitle{KMT-2019-BLG-1339L}
\shortauthors{Han et al.}

\begin{document}

\title{KMT-2019-BLG-1339L: an M Dwarf with a Giant Planet or a Companion
Near the Planet/Brown Dwarf Boundary}

\author{
Cheongho~Han$^{0001}$, 
Doeon~Kim$^{0001}$,
Andrzej~Udalski$^{0003,100}$, 
Andrew~Gould$^{0002,0004,0005,101}$,
\\
(Leading authors),\\
and \\
Michael~D.~Albrow$^{0007}$, 
Sun-Ju~Chung$^{0002,0008}$,  
Kyu-Ha~Hwang$^{0002}$, 
Youn~Kil~Jung$^{0002}$, 
Chung-Uk~Lee$^{0002,101}$, 
Yoon-Hyun~Ryu$^{0002}$, 
In-Gu~Shin$^{0002}$, 
Yossi~Shvartzvald$^{0009}$, 
Jennifer~C.~Yee$^{0010}$, 
Weicheng~Zang$^{0011}$,
Sang-Mok~Cha$^{0002,0012}$, 
Dong-Jin~Kim$^{0002}$, 
Hyoun-Woo~Kim$^{0002}$, 
Seung-Lee~Kim$^{0002,0008}$, 
Dong-Joo~Lee$^{0002}$, 
Yongseok~Lee$^{0002,0012}$, 
Byeong-Gon~Park$^{0002,0008}$, 
Richard~W.~Pogge$^{0005}$, 
\\
(The KMTNet Collaboration),\\
Przemek~Mr{\'o}z$^{0003,0018}$, 
Micha{\l}~K.~Szyma{\'n}ski$^{0003}$, 
Jan~Skowron$^{0003}$,
Rados{\l}aw~Poleski$^{0003}$, 
Igor~Soszy{\'n}ski$^{0003}$, 
Pawe{\l}~Pietrukowicz$^{0003}$,
Szymon~Koz{\l}owski$^{0003}$, 
Krzysztof~Ulaczyk$^{0015}$ \\
Krzysztof~A.~Rybicki$^{0003}$,
Patryk~Iwanek$^{0003}$,
Marcin~Wrona$^{0003}$,
Mariusz~Gromadzki$^{0003}$,\\
(The OGLE Collaboration) \\   
}

\email{cheongho@astroph.chungbuk.ac.kr}

\affil{$^{0001}$ Department of Physics, Chungbuk National University, Cheongju 28644, Republic of Korea} 
\affil{$^{0002}$ Korea Astronomy and Space Science Institute, Daejon 34055, Republic of Korea} 
\affil{$^{0003}$ Astronomical Observatory, University of Warsaw, Al.~Ujazdowskie 4, 00-478 Warszawa, Poland} 
\affil{$^{0004}$ Max Planck Institute for Astronomy, K\"onigstuhl 17, D-69117 Heidelberg, Germany} 
\affil{$^{0005}$ Department of Astronomy, Ohio State University, 140 W. 18th Ave., Columbus, OH 43210, USA} 
\affil{$^{0007}$ University of Canterbury, Department of Physics and Astronomy, Private Bag 4800, Christchurch 8020, New Zealand} 
\affil{$^{0008}$ Korea University of Science and Technology, 217 Gajeong-ro, Yuseong-gu, Daejeon, 34113, Republic of Korea} 
\affil{$^{0009}$ Department of Particle Physics and Astrophysics, Weizmann Institute of Science, Rehovot 76100, Israel }
\affil{$^{0010}$ Center for Astrophysics $|$ Harvard \& Smithsonian 60 Garden St., Cambridge, MA 02138, USA} 
\affil{$^{0011}$ Department of Astronomy and Tsinghua Centre for Astrophysics, Tsinghua University, Beijing 100084, China} 
\affil{$^{0012}$ School of Space Research, Kyung Hee University, Yongin, Kyeonggi 17104, Republic of Korea} 
\affil{$^{0015}$ Department of Physics, University of Warwick, Gibbet Hill Road, Coventry, CV4 7AL, UK} 
\affil{$^{0016}$ Department of Astronomy \& Space Science, Chungbuk National University, Cheongju 28644, Republic of Korea} 
\affil{$^{0018}$ Division of Physics, Mathematics, and Astronomy, California Institute of Technology, Pasadena, CA 91125, USA}
\altaffiltext{100}{OGLE Collaboration.}
\altaffiltext{101}{KMTNet Collaboration.}

\begin{abstract}
We analyze KMT-2019-BLG-1339, a microlensing event 
with an obvious but incompletely resolved brief anomaly feature around the peak of the light curve.  
Although the origin of the anomaly is identified to be a companion to the lens with a low mass ratio 
$q$, the interpretation is subject to two different degeneracy types.  The first type is the ambiguity 
in $\rho$, representing the angular source radius scaled to the angular radius of the Einstein ring, 
$\theta_{\rm E}$, and the other is the $s\leftrightarrow s^{-1}$ degeneracy.  The former type, 
`finite-source degeneracy', causes ambiguities in both $s$ and $q$, while the latter induces an 
ambiguity only in $s$.  Here $s$ denotes the separation (in units of $\theta_{\rm E}$) in projection 
between the lens components.  We estimate that the lens components have masses 
$(M_1, M_2)\sim (0.27^{+0.36}_{-0.15}~M_\odot, 11^{+16}_{-7}~M_{\rm J})$ and 
$\sim (0.48^{+0.40}_{-0.28}~M_\odot, 1.3^{+1.1}_{-0.7}~M_{\rm J})$ according to the two solutions 
subject to the finite-source degeneracy, indicating that the lens comprises an M dwarf and a companion 
with a mass around the planet/brown dwarf boundary or a Jovian-mass planet.  It is possible to lift 
the finite-source degeneracy by conducting future observations utilizing a high resolution instrument 
because the relative lens-source proper motion predicted by the solutions are widely different.
\end{abstract}


\section{Introduction}\label{sec:one}

A microlensing planet is, in general, detected through a short-lasting anomaly appearing in the 
lensing light curve of the host \citep{Mao1991}.  Due to the brief nature of a planetary signal, 
one is often confronted with cases in which the coverage of the signal is incomplete due to 
various causes such as insufficient cadence of observations, bad weather, time gap between 
observatories, etc.

Interpreting short-term microlensing signals with incomplete coverage can result in a degeneracy 
problem, in which multiple solutions with different combinations of lensing parameters can describe 
an observed anomaly, leading to multiple interpretations of the signal.  There have been reports 
of such cases caused by various types of degeneracy.  The first case was reported by \citet{Skowron2018} 
for OGLE-2017-BLG-0373 with a partially covered planetary signal.  From the analysis of the light 
curve, they found a pair of degenerate solutions, in which the folds of the planetary caustic located 
on the opposite sides with respect to the caustic center were swept by the source with nearly equal 
offsets from the caustic center.  The two solutions resulting from this ``caustic-chiral degeneracy'' 
have substantially different values of the planet/host mass ratio $q$, although they have similar 
planet-host separations (in projection) $s$ (scaled to the angular Einstein radius $\thetae$).  This 
mode of degeneracy was also identified by \citet{Hwang2018a}, when they analyzed KMT-2016-BLG-0212 
with a partially covered planetary signal.  The second case of a discrete degeneracy was identified 
from the analysis of OGLE-2017-BLG-0173 \citep{Hwang2018b}, for which the anomaly was insufficiently 
covered with a gap in the data.  This  ``Hollywood degeneracy'' yielded two solutions, in which the 
source fully encompassed the planetary caustic according to one solution, and the source surrounded 
only one caustic side according to the other solution.  This degeneracy causes an ambiguity in 
determining $q$.  The third case, found from OGLE-2018-BLG-0740 with a partially covered planetary 
signal, was reported by \citet{Han2019}.  For this event, the degenerate solutions yielded different 
values of $\rho$, representing the angular radius of the source $\theta_*$ normalized to $\thetae$ 
(normalized source radius), and this caused ambiguities in both $s$ and $q$.   Considering that 
planetary signals for an important fraction of microlensing events would be detected with incomplete 
coverage, it is important to identify various types of degeneracies and investigate their origins to 
correctly interpret planetary signals in future analyses.

Here we analyze the partially covered short-term anomaly feature that appears in the lensing light curve of 
KMT-2019-BLG-1339.  We find that interpreting the anomaly is subject to two different types of 
degeneracy, and we investigate the origins of the degeneracies.

The organization of the paper is as follows.  The data acquired from observing the lensing event 
are addressed in Section~\ref{sec:two}.  In Section~\ref{sec:three}, we mention the procedure of 
modeling conducted to interpret the observed anomaly.  We also mention the types of 
degeneracy identified from modeling.  We estimate the $\thetae$ values corresponding to the 
degenerate solutions in Section~\ref{sec:four}.  Estimation of the lens masses and distances for 
the degenerate solutions is provided in Section~\ref{sec:five}.  In Section~\ref{sec:six},  we 
suggest a method to lift the identified degeneracy.  Summary of the findings and conclusion 
are given in Section~\ref{sec:seven}.

\section{Observations}\label{sec:two}

The source star of the lensing event 
KMT-2019-BLG-1339/OGLE-2019-BLG-1019 lies in the Galactic 
bulge field.  The coordinates of the source are $({\rm R.A.}, {\rm decl.})_{\rm J2000}\equiv 
(17:42:58.42, -25:34:26.1)$, corresponding to  
$(l, b)=(2^\circ\hskip-2pt .559, 2^\circ\hskip-2pt .260)$. 
The apparent brightness of the source remained constant before lensing with an $I$-band baseline magnitude
of $I\sim 20.4$.

\begin{figure}
\includegraphics[width=\columnwidth]{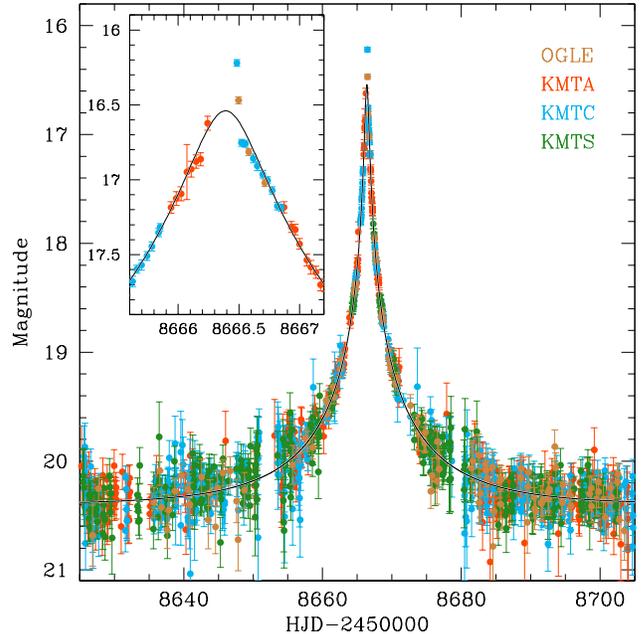}
\caption{
Photometric data around KMT-2019-BLG-1339.
The curve plotted over the data points is the 1L1S model. 
The zoom of the peak region is shown in the inset.  
\smallskip
}
\label{fig:one}
\end{figure}

The Korea Microlensing Telescope Network \citep[KMTNet:][]{Kim2016} experiment first detected 
the event on 2019-06-26, which corresponds to ${\rm HJD}^\prime\equiv {\rm HJD}-2450000=8660$, 
using the alert-finder system \citep{Kim2018}.  At the time of finding, the source was brighter 
than the baseline magnitude by $\Delta I \sim 0.86$~magnitude.  Five days later, the event was 
independently found by another lensing survey of the Optical Gravitational Lensing Experiment 
\cite[OGLE:][]{Udalski2015}. The event was designated as KMT-2019-BLG-1339 and OGLE-2019-BLG-1019 
by the individual surveys. Hereafter, we use KMT-2019-BLG-1339 as a representative name of the 
event.  The KMTNet survey utilizes three identical telescopes, each with a 1.6~m aperture and  
mounted with a camera having a $4~{\rm deg}^2$ field of view.  The telescopes are located in 
three different continents: the South African Astronomical Observatory in South Africa (KMTS), 
the Cerro Tololo Inter-American Observatory in Chile (KMTC), and the Siding Spring Observatory 
in Australia (KMTA).  The OGLE survey utilized the Warsaw telescope, with a 1.3~m aperture, at 
the Las Campanas Observatory in Chile, and the camera mounted on the telescope has a 
$1.4~{\rm deg}^2$ field of view.  The source was located in the KMTNet BLG19 and OGLE BLG652.26 
fields, toward which observations by the individual surveys were conducted with $\sim 1~{\rm hr}$ 
and $\sim 2~{\rm hr}$ cadences, respectively. For both surveys, observations were conducted 
primarily in the $I$ band, and observations in the $V$ band were done to obtain a subset of data 
to measure the source color.  The data used in the analysis lie in the time range 
of $8620\lesssim {\rm HJD}^\prime \lesssim 8720$.

Reduction of data is done with the photometry pipelines of the KMTNet \citep{Albrow2009} and OGLE 
\citep{Wozniak2000} surveys.  These pipelines commonly employ the difference image analysis (DIA) 
algorithm \citep{Alard1998, Tomaney1996}, that is developed for optimal crowded field photometry.  
For a subset of KMTC $I$- and $V$-band images taken around the lensing magnifications, we 
carry out extra photometry with the use of the pyDIA software \citep{Albrow2017} to measure the 
source color.  
The ranges of the pyDIA data sets are $8566.9 \leq  {\rm HJD}^\prime \leq 8728.5$ and 
$8642.7 \leq  {\rm HJD}^\prime \leq 8713.5$ for the $I$ and $V$-band data sets, respectively. 
We note that the pyDIA photometry measures the flux itself,  while the DIA photometry measures 
the difference in flux from the baseline.  Because the pyDIA photometry is affected by blended light, 
the photometry quality is poorer than that of the DIA photometry.  Nevertheless, 
data sets processed by the pyDIA photometry 
are needed to estimate the apparent magnitudes of the lensing event and the color of the source
star.  Detailed process of the source color estimation is discussed in Section~\ref{sec:four}.

We reevaluate the errors bars of data from the photometry pipelines.
Following the recipe addressed in \citet{Yee2012}, this process is done by 
\begin{equation}
\sigma=k(\sigma_0^2+\sigma_{\rm min}^2)^{1/2}.
\label{eq1}
\end{equation}
Here $\sigma_0$ denotes the error bar estimated from the pipeline, $\sigma_{\rm min}$ is the 
scatter of data, and $k$ is a factor used to make $\chi^2$ per degree of freedom become unity.  
Table~\ref{table:one} shows the values of $k$ and $\sigma_{\rm min}$ together with the numbers of 
data points, $N_{\rm data}$, in the individual data sets.

Figure~\ref{fig:one} shows the photometry data around the time of the lensing event.  Different 
colors are used to designate the telescopes used for the data acquisition.  The solid curve plotted 
over the data points is the model found under a standard single-lens single-source (1L1S) 
interpretation.  According to the 1L1S model, the event is magnified with a moderately high 
magnification of $A_{\rm peak}\sim 62$ at the peak, and the event timescale is $\te\sim 17.6$ days.  
We check the feasibility of measuring the annual microlens parallax $\pie$ \citep{Gould1992}, but 
$\pie$ cannot be securely estimated mostly because of the short event timescale relative to the 
orbital period of the Earth.

\begin{deluxetable}{lccc}
\tablecaption{Numbers of data points\label{table:one}}
\tablewidth{240pt}
\tablehead{
\multicolumn{1}{c}{Data set}              &
\multicolumn{1}{c}{$k$}                   &
\multicolumn{1}{c}{$\sigma_{\rm min}$}    &
\multicolumn{1}{c}{$N_{\rm data}$}        
}
\startdata                                              
KMTA       &  1.068 &  0.030  &   355    \\
KMTC       &  1.054 &  0.020  &   569    \\
KMTS       &  1.034 &  0.040  &   426    \\
OGLE       &  1.082 &  0.020  &   125    
\enddata                            
\end{deluxetable}

\section{Light Curve Analysis}\label{sec:three}

Although the light curve of the event seemingly looks like that of a 1L1S event, a close 
inspection reveals that the peak part of the light curve exhibits small but noticeable 
deviations from the model.  See the inset of Figure~\ref{fig:one} showing the zoom of the 
peak region.

\begin{figure}
\includegraphics[width=\columnwidth]{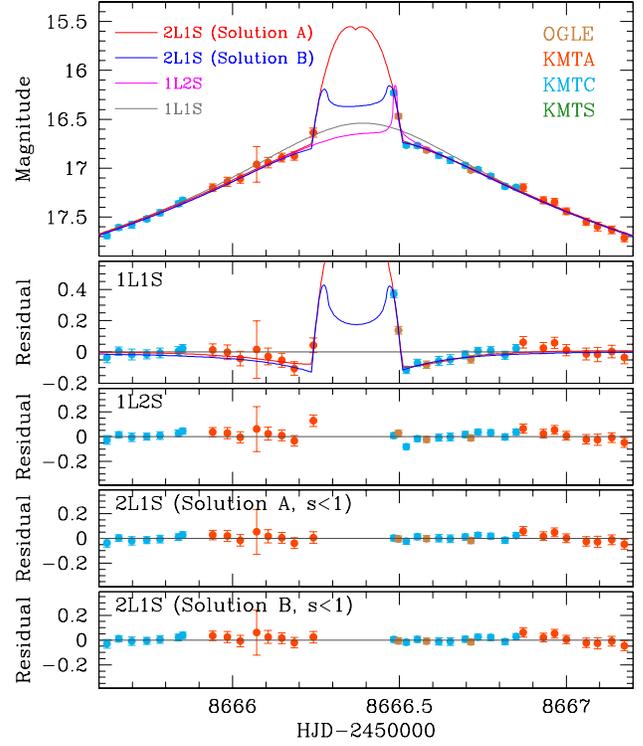}
\caption{
Model curves under various interpretations of the lens system, including 1L1S , 1L2S, and 
2L1S models.  For the 2L1S interpretation, we present two models, solution ``A''  (with 
$s<1.0$) and ``B'' ($s<1.0$), resulting from the discrete degeneracy. The residuals from the 
tested models are shown in the lower panels.  The two curves drawn on the 1L1S residuals 
(second panel) are the 
differences between the 2L1S (red for solution A and blue for solution B) and 1L1S models.
\smallskip
}
\label{fig:two}
\end{figure}

Figure~\ref{fig:two} shows the residuals from the 1L1S model around the peak.  
The residuals exhibit the following characteristics. First, the three data points at 
${\rm HJD}^\prime=8666.242$ (KMTA), 8666.483 (KMTC), and 8666.497 (OGLE) exhibit discontinuous 
deviations from the 1L1S model. Second, the KMTA data points just before the peak (in the region 
$8666.0\lesssim {\rm HJD}^\prime \lesssim 8666.2$) and the OGLE and KMTC data points just after 
the peak (in the region $8666.5\lesssim {\rm HJD}^\prime \lesssim  8666.7$) exhibit continuous 
negative deviations from the 1L1S model. We note that the coverage of the major part of the 
anomaly is incomplete due to the $\sim 5.8$~hr time gap between the last point of the KMTA data 
and the first point of the KMTC data taken on 2019-07-01.  The gap corresponded to a night time 
in Africa, but observations by KMTS could not be conducted due to the bad weather.

The characteristics of the deviations from the 1L1S model suggest that there may exist a caustic 
in the central magnification zone induced by a companion to the lens.  With such a caustic, the 
three data points exhibiting discontinuous deviations would be explained by the caustic crossings 
of the source, and the data points with smooth negative deviations before and after the peak 
would be explained by the negative excess magnification in the regions immediately outside the 
fold of the caustic.  If this interpretation is correct, the caustic should be very small because 
the time gap between the two successive caustic crossings is very short.

We check the presence of a central caustic by modeling the light curve under the binary lens 
interpretation (2L1S model).  A 1L1S light curve is defined by three parameters, which are the 
time of the lens-source minimum separation, $t_0$, the impact parameter, $u_0$, and  the event 
timescale, $t_{\rm E}$.  A 2L1S modeling demands extra parameters of $(s, q, \alpha)$.  Here 
$\alpha$ denotes the source trajectory angle, which is defined as the angle between the source 
trajectory and the line connecting the binary lens components.  Because the three discontinuous 
points are believed to lie on the caustic-crossing parts, during which lensing magnifications 
experience finite-source effects.  To account for these effects, we conduct modeling with the 
inclusion of an extra parameter $\rho\equiv \theta_*/\thetae$ (normalized source radius).  Here 
$\theta_*$ represents the angular size of the source radius.  We compute finite-source magnifications 
using the method of \citet{Dong2006}, which utilizes the ray-shooting algorithm.  We take the 
limb-darkening effect into consideration in computing finite magnifications.  We choose the 
limb-darkening coefficients considering the source type, which is a late F-type main-sequence 
star.  Details of the source type determination are mentioned in Section~\ref{sec:four}.  We 
assume that the surface brightness varies as $S\propto 1-\Gamma_\lambda (1-3 \cos\psi/2)$, in 
which $\Gamma$ denotes the limb-darkening coefficient and $\psi$ is the angle between the radial 
direction from the source center and the line of sight toward the source center.  We adopt 
$\Gamma_I=0.37$ and $\Gamma_V=0.52$ from the \citet{Claret2000} catalog.

Implementing light curve modeling is done in two rounds.  In the first-round modeling, 
we find the parameters $s$ and $q$ using a grid-search approach, while we look for the other 
parameters utilizing a downhill simplex method based on the Markov Chain Monte Carlo (MCMC) 
algorithm.  In the second-round modeling, we find local minima by inspecting $\Delta\chi^2$ 
map on the grid parameter, $s$ and $q$, space.  Each local minima is then further refined from 
an additional modeling, in this time, by letting all parameters vary.  This two-step procedure 
is useful in identifying degenerate solutions, if they exist.

\begin{figure}
\includegraphics[width=\columnwidth]{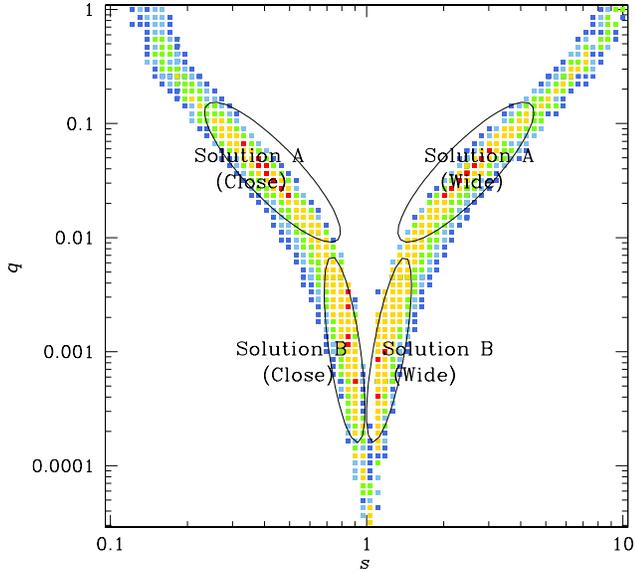}
\caption{
Four local minima in the $\Delta\chi^2$ map on the parameter plane of the binary separation $s$ 
and mass ratio $q$.  The colors of points designate regions with $<1n\sigma$ (red), $<2n\sigma$ 
(yellow), $<3n\sigma$ (green), and $<4n\sigma$ (blue), where $n=3$.
\smallskip
}
\label{fig:three}
\end{figure}

The 2L1S modeling yields four degenerate local solutions.  For visual presentation of the local 
solutions, we mark the individual degenerate solutions in the $s$--$q$ plane in Figure~\ref{fig:three}.  
From the inspection of the map, it is found that the mass ratios of one pair of the solutions are 
$q\sim 4\times 10^{-2}$ and those of the other pair are $q\sim 2.5\times 10^{-3}$.  Hereafter, we 
refer to the solutions with $q\sim 4\times 10^{-2}$ and $q\sim 2.5\times 10^{-3}$ as solutions 
``A'' and ``B'', respectively.  For each pair, we identify another pair of solutions: one with 
$s<1$ (close solution) and the other with $s>1$ (wide solution).  The latter degeneracy is caused 
by the close/wide degeneracy \citep{Griest1998, Dominik1999, An2005}.  With $\chi^2$ differences 
among the solutions being merely $\Delta\chi^2\sim 1.2$, we find that resolving the degeneracies 
is difficult based on the photometric data alone.

It is found that the 2L1S solutions well describe the observed light curve including the peak 
part exhibiting deviations from the 1L1S model.  The 2L1S models provides a better fit than the 
1L1S model by $\Delta\chi^2\sim 290$.  The model curves of the solutions A (with $s<1$) and B ($s<1$) 
are shown in the top panel of Figure~\ref{fig:two}, and the residuals from the individual models 
are presented in the bottom two panels.  In the second panel, we present the difference between the 
2L1S and 1L1S models: red curve for the solution A (with $s<1$) and blue curve for the solution B ($s<1$).  
For both solutions A and B, the deviations from the 1L1S solution are explained by the caustic crossings, 
as expected.  However, it is found that the model curves of the two degenerate solutions are substantially
different in the region between the times of the caustic crossings.  The degeneracy could have been 
resolved if there existed a few data points between the times of the caustic crossings, implying that 
the degeneracy is accidentally caused by the gap in the data.

\begin{figure}
\includegraphics[width=\columnwidth]{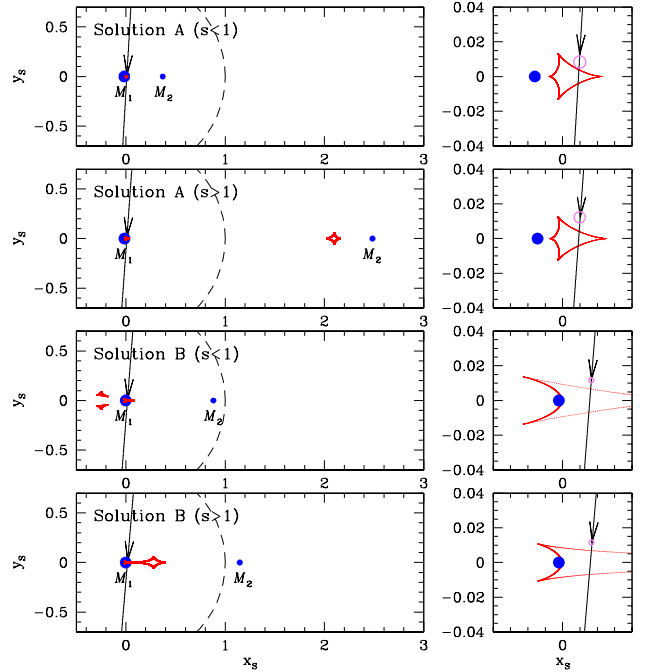}
\caption{
Configurations of the lens and source for the four local solutions from the 2L1S modeling.  In each 
panel, the two blue dots, labeled by $M_1$ and $M_2$, denote the lens positions and the cuspy closed 
figures are the caustics.  The solid line and the arrow on the line indicate the trajectory and the 
direction of the source motion, respectively.  The small pink circle on the source trajectory is drawn 
to compare the source size with the caustic size.  The dashed circle is the Einstein ring.  
\smallskip
}
\label{fig:four}
\end{figure}

\begin{deluxetable*}{lccccc}
\tablecaption{Lensing parameters of various models\label{table:two}}
\tablewidth{480pt}
\tablehead{
\multicolumn{1}{c}{Parameter}          &
\multicolumn{1}{c}{1L2S}    &
\multicolumn{2}{c}{2L1S (Solution A)}  &
\multicolumn{2}{c}{2L1S (Solution B)}  \\   
\multicolumn{1}{c}{}                   &
\multicolumn{1}{c}{}                   &
\multicolumn{1}{c}{Close}              &
\multicolumn{1}{c}{Wide}               &
\multicolumn{1}{c}{Close}              &
\multicolumn{1}{c}{Wide}               
}
\startdata                                              
$\chi^2$                     & 1499.5                       &  1474.5                &  1474.3                 &  1473.3                 &  1473.8               \\
$t_0$ (${\rm HJD}^\prime$)   & $8666.395 \pm   0.004$       &  $8666.378 \pm 0.004$  &  $8666.379 \pm 0.004$   &  $8666.391 \pm 0.003 $  &  $8666.393 \pm 0.003$ \\
$u_0$ ($10^{-2}$)            & $2.002 \pm  0.129$           &  $0.913 \pm 0.081   $  &  $0.922 \pm 0.083   $   &  $1.560 \pm 0.089    $  &  $1.593 \pm 0.096   $ \\
$t_{\rm E}$ (days)           & $16.43 \pm  0.67$            &  $17.23 \pm 0.84    $  &  $17.66 \pm 0.90    $   &  $16.70 \pm 0.68     $  &  $16.45 \pm 0.70    $ \\
$s$                          & --                           &  $0.39 \pm 0.05     $  &  $2.50 \pm 0.41     $   &  $0.88 \pm 0.03      $  &  $1.15 \pm 0.04     $ \\
$q$ ($10^{-3}$)              & --                           &  $43.03 \pm 13.94   $  &  $37.48 \pm 15.22   $   &  $2.47 \pm 0.66      $  &  $2.50 \pm 0.77     $ \\
$\alpha$ (rad)               & --                           &  $4.771 \pm 0.019   $  &  $4.784 \pm 0.016   $   &  $4.790 \pm 0.014    $  &  $4.797 \pm 0.015   $ \\
$\rho$ ($10^{-3}$)           & --                           &  $3.49 \pm 0.61     $  &  $3.20 \pm 0.54     $   &  $1.42 \pm 0.32      $  &  $1.38 \pm 0.30     $ \\
$t_{0,2}$                    & $8666.487 \pm 0.003$         &  --                    &  --                     &  --                     &  --                   \\  
$u_{0,2}$ ($10^{-2}$)        & $0.000 \pm 0.022$            &  --                    &  --                     &  --                     &  --                   \\  
$\rho_2$ ($10^{-3}$)         & $0.39 \pm 0.21$              &  --                    &  --                     &  --                     &  --                   \\  
$q_F$                        & $0.006 \pm0.002$             &  --                    &  --                     &  --                     &  --                   \\  
$f_{\rm s,OGLE}$             & $0.068 \pm 0.003   $         &  $0.063 \pm 0.003   $  &  $0.063 \pm 0.004   $   &  $0.066 \pm 0.003    $  &  $0.067 \pm 0.003   $ \\
$f_{\rm b,OGLE}$             & $0.042 \pm 0.003   $         &  $0.047 \pm 0.003   $  &  $0.046 \pm 0.003   $   &  $0.044 \pm 0.003    $  &  $0.043 \pm 0.003   $
\enddata                            
\tablecomments{
${\rm HJD}^\prime \equiv {\rm HJD}-2450000$.
\bigskip
}
\end{deluxetable*}


The lensing parameters and the $\chi^2$ values of the fits for the individual 2L1S solutions 
are listed in Table~\ref{table:two}.  The mass ratios of the B solutions, $q\sim 2.5\times 10^{-3}$, 
indicate that the companion to the lens has a planetary mass.  For the A solutions with 
$q\sim 4\times 10^{-2}$, on the other hand, the nature of the lens companion is uncertain just 
based the mass ratio because the mass could be below or above the lower mass limit of brown dwarfs 
(BDs), $\sim 13~M_{\rm J}$ \citep{Boss2007}, depending on the mass of the primary.  The $f_{s,I}$ 
and $f_{b,I}$ listed in the table denote the values of the $I$-band flux for the source and blend, 
respectively.  Following the OGLE photometry system, the flux is set to be unity for a star with 
$I=18$.  Besides the mass ratios, we point out that the normalized source radii estimated from the 
solutions A and B are substantially different: $\rho\sim 3.3\times 10^{-3}$ (solutions A) and 
$\rho\sim 1.4\times 10^{-3}$ (solutions B). As a result, the model curve of the solution B exhibits 
a well-defined ``U''-shape trough feature between the caustic crossings, while the feature in the 
model curve of the solution A is smeared out by severe finite-source effects.

The configurations of the source and lens for the individual 2L1S solutions are shown in 
Figure~\ref{fig:four}.  For each solution, the left panel is presented to show the locations of 
both lens components (blue dots tagged by $M_1$ and $M_2$), and the right panel is given to show 
the central magnification region.  For solutions~B, the lower-mass lens component lies near the 
Einstein ring (dashed circle centered at the origin).  For solutions~A, on the other hand, the 
separation of the lens companion from the Einstein ring is relatively big.  It is found that the 
central caustics for each close-wide pair of the solutions appear to be similar to each other, 
and this results in the close/wide degeneracy.  In contrast, the caustics between the solutions 
A and B appear to be substantially different.  The solid line and the arrow on the line indicate 
the trajectory and the direction of the source motion, respectively.  It shows that the anomaly 
arises by the source passage through the central caustic almost at a right angle.

The mode of the degeneracy between the solutions A and B is similar to the degeneracy mode
identified for OGLE-2018-BLG-0740 \citep{Han2019}.  The similarity between the two events is 
that the coverage of the anomaly is incomplete, and this causes the ambiguity in $\rho$.  For 
OGLE-2018-BLG-0740, the ``U''-shape trough feature is covered, but the caustic-crossing parts 
are poorly covered.  For KMT-2019-BLG-1339, on the other hand, there are three data points, one 
in rising and two in falling parts, in the caustic-crossing parts, but the ``U''-shape trough 
feature is not covered.  This suggests that the finite-source degeneracy can arise when the 
coverage of an anomaly is incomplete.

A short-duration anomaly can also be produced by a subset of binary-source (1L2S) events
\citep{Gaudi1998, Gaudi2004, Shin2019}, and thus we conduct an additional modeling with the 
1L2S interpretation.  Similar to the 2L1S case, a 1L2S modeling requires to include extra 
parameters in addition to those of the 1L1S modeling.  According to the parameterization of 
\citet{Hwang2013}, these extra parameters are $t_{0,2}$, $u_{0,2}$, $\rho_2$, and $q_F$, which 
represent the time of the closest lens approach to the source companion, the impact parameter 
of the source companion motion, the normalized radius of the source companion, and the flux 
ratio between the binary source stars, respectively.

The best-fit 1L2S model and its residuals are shown in Figure~\ref{fig:two}.  The lensing 
parameters of the model are listed in Table~\ref{table:two}.  We note that the lens passes 
over the surface of the source companion according to the model, but the lens does not 
transverse the primary source.  As a result, the value of $\rho_2$ is presented, while the 
value of $\rho$ is not given.  From the comparison of the fits, it is found that the 1L2S 
gives a better fit than the 1L1S model by $\Delta\chi^2\sim 264$, but the fit is worse than 
the 2L1S model by $\Delta\chi^2\sim 26$.  Therefore, we conclude that the central perturbation 
is caused not by a source companion but by a lens companion.

\begin{figure}
\includegraphics[width=\columnwidth]{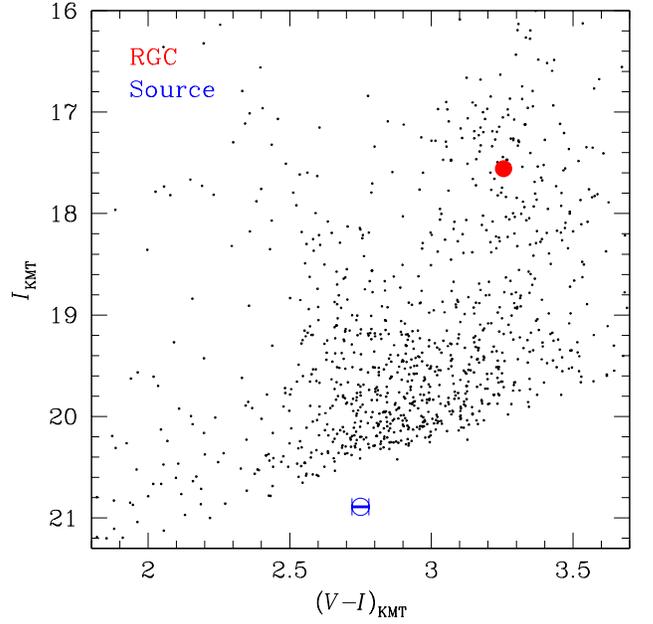}
\caption{
Color-magnitude diagram of stars around the source of KMT-2019-BLG-1339.
The blue empty dot indicates the source position and
the red filled dot denotes the centroid of red giant clump (RGC).
\smallskip
}
\label{fig:five}
\end{figure}

\section{Angular Einstein Radius}\label{sec:four}

The solutions A and B result in significantly different values of the Einstein radius.  
This is because these solutions yield substantially different values of $\rho$, 
from which the value of $\thetae$ is determined, i.e., $\thetae=\theta_*/\rho$.

For the $\thetae$ determination, we first estimate $\theta_*$.  We estimate $\theta_*$ using 
the dereddened color and brightness, $(V-I,I)_0$, of the source.  Following the recipe of 
\citet{Yoo2004}, we determine the positions of the source and the centroid of red giant clump 
(RGC) in the color-magnitude diagram (CMD), measure the  offset between the source and RGC 
centroid, and then estimate $(V-I,I)_0$ of the source based on the known values of the dereddened 
color and magnitude of the RGC centroid, $(V-I, I)_{\rm RGC,0}$.

Figure~\ref{fig:five} shows the instrumental CMD of stars in the vicinity of the source, and the 
locations of the source, at $(V-I, I)$, and the RGC centroid, at $(V-I, I)_{\rm RGC}$.  
The CMD construction and $(V-I, I)$ estimation are based on the KMTC data set reduced using 
the same pyDIA photometry.  
We note that the location of the blend is not marked because the color 
cannot be determined due to the poor $V$-band photometry.
The measured values of the color and magnitude are 
$(V-I,I)=(2.75\pm 0.04, 20.89\pm 0.01)$ and $(V-I,I)_{\rm RGC}=(3.26, 17.56)$ for the source and 
RGC centroid, respectively.  We calibrate the color and magnitude of the source star using its 
offsets from the RGC centroid, $\Delta(V-I, I)$, by using the relation 
\begin{equation} 
(V-I, I)_0= (V-I, I)_{\rm RGC,0}+ \Delta(V-I, I),  
\label{eq2} 
\end{equation} 
We adopt $(V-I, I)_{\rm RGC,0}=(1.06, 14.32)$ from \citet{Bensby2013} and \citet{Nataf2013}.
From the process, we find $(V-I,I)_0=(0.55\pm 0.04, 17.65\pm 0.01)$.  These values point out that 
the lensing event occurred on a bulge main sequence with a late F spectral type.  After this 
calibration process, we apply the color-color relation of \citet{Bessell1988} to derive $V-K$ 
color, and then use the the \citet{Kervella2004} relation between color and surface-brightness 
to derive $\theta_*$.  From this process, the angular radius of the source is estimated as 
\begin{equation}
\theta_* = 0.78 \pm 0.06~\mu{\rm as}.
\label{eq3}
\end{equation}

\begin{deluxetable}{lcc}
\tablecaption{Angular Einstein radius and relative proper motion\label{table:three}}
\tablewidth{240pt}
\tablehead{
\multicolumn{1}{c}{Quantity}  &
\multicolumn{1}{c}{Solution A}  &
\multicolumn{1}{c}{Solution B}           
}
\startdata                                              
$\thetae$ (mas)            &   $0.22 \pm 0.04    $   &   $0.55  \pm 0.13   $ \\
$\mu$ (mas yr$^{-1}$)      &   $4.76 \pm 0.92    $   &   $12.08 \pm 2.89   $ 
\enddata                            
\end{deluxetable}

\begin{deluxetable*}{lcccc}
\tablecaption{Physical lens parameters\label{table:four}}
\tablewidth{480pt}
\tablehead{
\multicolumn{1}{c}{Parameter}  &
\multicolumn{2}{c}{Solution A}  &
\multicolumn{2}{c}{Solution B}  \\   
\multicolumn{1}{c}{}  &
\multicolumn{1}{c}{Close}  &
\multicolumn{1}{c}{Wide}  &
\multicolumn{1}{c}{Close}  &
\multicolumn{1}{c}{Wide}           
}
\startdata                                              
$M_1$  $(M_\odot)$     & $0.27^{+0.36}_{-0.15}$   &  $0.26^{+0.35}_{-0.14}$   &   $0.48^{+0.40}_{-0.28}$    &  $0.49^{+0.40}_{-0.28}$ \\
$M_2$ $(M_{\rm J})$    & $12.2^{+16.1}_{-6.7} $   &  $10.7^{+14.0}_{-5.8} $   &   $1.25^{+1.04}_{-0.73}$    &  $1.27^{+1.05}_{-0.74}$ \\
$D_{\rm L}$ (kpc)      & $7.15^{+1.06}_{-1.25}$   &  $7.21^{+1.04}_{-1.22}$   &   $6.12^{+1.26}_{-1.60}$    &  $6.00^{+1.26}_{-1.64}$ \\
$d_\perp$   (au)       & $0.67^{+0.77}_{-0.55}$   &  $4.30^{+4.94}_{-3.55}$   &   $2.15^{+2.59}_{-1.58}$    &  $2.80^{+3.38}_{-2.07}$
\enddata                            
\end{deluxetable*}

With the measured $\theta_*$, we estimate the values of $\thetae$ and $\mu$ by
\begin{equation}
\thetae={\theta_*\over \rho};\qquad 
\mu = {\thetae\over t_{\rm E}}
\label{eq4}
\end{equation}
respectively.  The values of $\thetae$ and $\mu$ for the solutions A and B are listed in 
Table~\ref{table:three}.  To be noted is that the Einstein radius for the solution B, 
$\sim 0.55$~mas, is bigger than the value estimated from the solution A, $\sim 0.22$~mas, by 
a factor $\sim 2.5$.  The difference in $\thetae$ originates from the difference in the values 
of $\rho$ estimated from the individual solutions.  For the same reason, the $\mu$ value of the 
solution~A, $\sim 4.8~{\rm mas}~{\rm yr}^{-1}$, is smaller than the value estimated from the 
solution~B, $\sim 12.1~{\rm mas}~{\rm yr}^{-1}$, by a similar factor.

\section{Physical Lens Parameters}\label{sec:five}

The lens mass $M$ and distance $D_{\rm L}$ are uniquely determined by measuring both $\thetae$ 
and $\pie$, i.e., 
\begin{equation}
M={\thetae\over \kappa\pie};\qquad
D_{\rm L}= {{\rm au}\over \pie\thetae + \pi_{\rm S}}.
\label{eq5}
\end{equation}
Here $\kappa=4G/(c^2{\rm au})$ and $\pi_{\rm S}$ represents the parallax to the source, 
i.e., $\pi_{\rm S} ={\rm au}/D_{\rm S}$. For KMT-2019-BLG-1339, $\thetae$ is measured with a
 two-fold degeneracy, but $\pie$ cannot be measured.  For the estimations $M$ and $D_{\rm L}$, 
we, therefore, conduct a Bayesian analysis based on the measured values of $t_{\rm E}$ and 
$\thetae$.

In the Bayesian analysis, we conduct a Monte Carlo simulation to produce numerous ($2\times 10^7$) 
artificial lensing events.  For the individual events, we derive the physical parameters of lenses 
(including the lens mass $M$, distance $D_{\rm L}$, and transverse lens-source speed $v$) from 
priors.  Lens masses are derived from the mass function of \citet{Chabrier2003} for stellar lenses 
and from the mass function of \citet{Gould2000} for remnant objects.  Locations and motion of the 
lens and source are derived from the \citet{Han2003} and \citet{Han1995} models, respectively.
For the individual 
events produced by the simulation, we compute the timescales, $t_{{\rm E},i}=D_{\rm L}\thetae/v$, and 
Einstein radii, $\theta_{{\rm E},i}=(\kappa M\pi_{\rm rel})^{1/2}$.  Here 
$\pi_{\rm rel}={\rm au}(D_{\rm L}^{-1}-D_{\rm S}^{-1})$ denotes the relative lens-source parallax.  
We then construct the posteriors for $M$ and $D_{\rm L}$ by imposing a weight factor $\exp(-\Delta\chi^2/2)$, 
where $\Delta\chi^2=[(t_{{\rm E},i}-t_{\rm E})/\sigma(t_{\rm E})]^2 + 
[(\theta_{{\rm E},i}-\theta_{\rm E})/\sigma(\theta_{\rm E})]^2$ and 
$[t_{\rm E}\pm \sigma(t_{\rm E}), \theta_{\rm E}\pm \sigma(\theta_{\rm E})]$ represent the measured 
values of $t_{\rm E}$ and $\theta_{\rm E}$, respectively.

\begin{figure}
\includegraphics[width=\columnwidth]{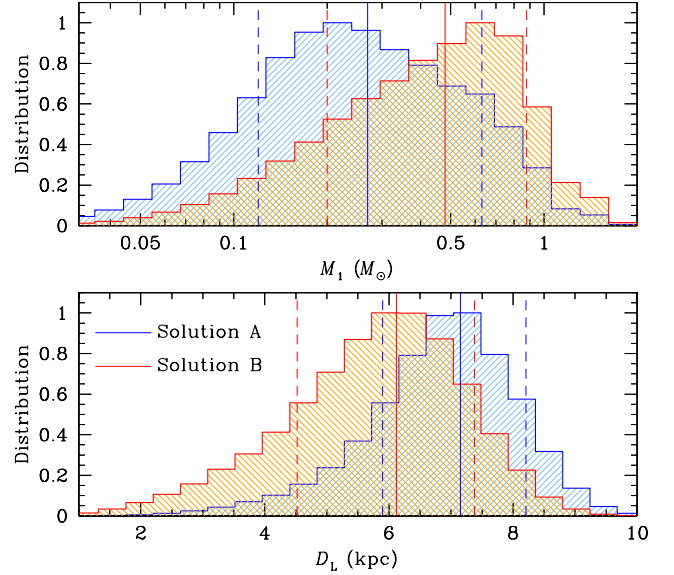}
\caption{
Posteriors for the primary lens mass, $M_1$, and the distance to the lens, $D_{\rm L}$. In each 
panel, the blue and red distributions are obtained based on the solutions A and B, respectively.  
For each posterior, the vertical solid line indicates the median and the two dashed lines (together 
with the line with arrows) represents $1\sigma$ range, estimated by the 16\% and 84\% of the 
distribution.
\smallskip
}
\label{fig:six}
\end{figure}

Figure~\ref{fig:six} shows the Bayesian posteriors for $M_1=M/(1+q)$ (mass of the primary, upper 
panel) and $D_{\rm L}$ (lower panel).  The blue and red curves  in each panel are the posteriors 
corresponding to the solutions A and B, respectively.  The estimated masses of the lens 
components, $M_1$ and $M_2$, distances, and  projected separations between the lens components, 
$d_\perp=sD_{\rm L}\thetae$, for the four degenerate solutions are listed in Table~\ref{table:four}.  
The presented values are the medians of the probability distributions and the uncertainties 
correspond to 16\% and 84\% of the distributions.  It is found that the primary lens has a mass
\begin{equation}
M_1 \sim 
\begin{cases}
0.27^{+0.36}_{-0.15}~M_\odot      & \text{for solution A},  \\
0.48^{+0.40}_{-0.28}~M_\odot      & \text{for solution B}.   
\end{cases}
\label{eq6}
\end{equation}
According to the median values of the individual solutions, the primary lens is a middle (for 
solution A) and an early (solution B) M dwarf.  However, the estimated masses from the two 
solutions overlap in a wide range because the uncertainties of the mass estimation 
are substantially bigger than the difference between the masses.  See the upper panel of 
Figure~\ref{fig:six}.
The mass of the lens companion is
\begin{equation}
M_2 \sim 
\begin{cases}
11^{+16}_{-7}~M_{\rm J}        & \text{for solution A},  \\
1.3^{+1.1}_{-0.7}~M_{\rm J}    & \text{for solution B}.   
\end{cases}
\label{eq7}
\end{equation}
The lens companion has a mass in the planetary regime
according to the solution B, while the mass of the 
companion lies at around the BD/planet boundary according to the solution A.  The estimated 
distance to the lens is
\begin{equation}
D_{\rm L} \sim 
\begin{cases}
7.2^{+1.1}_{-1.3}~{\rm kpc}      & \text{for solution A},  \\
6.1^{+1.3}_{-1.6}~{\rm kpc}      & \text{for solution B}.  
\end{cases}
\label{eq8}
\end{equation} Similar to the lens mass, the estimated distances from the two solutions overlap 
in a wide range, as shown in Figure~\ref{fig:six}.  For the solution A, in which the anomaly is 
produced by a low-mass companion with a binary separation considerably greater or smaller than 
unity, the projected separation greatly varies depending on the close/wide solution, with 
$d_\perp\sim 0.7$~au for the close solutions and $\sim 4.3$~au for the wide solution.  For 
solution B, in contrast, the difference in the projected separations between the close, 
$d_\perp\sim 2.2$~au, and wide solutions, $\sim 2.8$~au, is relatively small.

\section{Resolving Degeneracies}\label{sec:six}

We point out that the degeneracy between the solutions A and B can be lifted from future 
observations using high-resolution instrument.  These observations would enable one to directly 
measure the relative proper motion by resolving the lens and source, e.g., MACHO~LMC-5 
\citep{Alcock2001}, OGLE-2005-BLG-169 \citep{Bennett2015, Batista2015}, OGLE-2005-BLG-071 
\citep{Bennett2020}, and MOA-2013-BLG-220 \citep{Vandorou2019}.  Then, the degeneracy can be 
lifted because the two sets of solutions have substantially different values of the relative 
proper motion: $\mu\sim 4.8~{\rm mas}~{\rm yr}^{-1}$ for solution~A and 
$\sim 12.1~{\rm mas}~{\rm yr}^{-1}$ for solution~B.

The prospects for early lens-source resolution are more favorable for solution B.  This is because 
the solution yields a substantially higher relative proper motion than the solution~A.   For 
OGLE-2005-BLG-169, the lens and source could be resolved from Keck AO observations conducted when 
the lens-source separation was $\sim 50$~mas \citep{Batista2015}.  With the imposition of the same 
criterion, KMT-2019-BLG-1339L, where ``L'' denotes the lens, can be resolved from the source if Keck 
AO observation is conducted $\sim 4.1$~years after the event (in the second half of 2023).  Considering 
that the lens of KMT-2019-BLG-1339, 
with an expected $H$-band magnitude of $H\sim 20.1$ according to the solution B, 
is fainter 
than that of OGLE-2005-BLG-169, with $H\sim 17.9$, the lens resolution would take somewhat longer than 
$\sim 4.1$~years due to the low signal from the lens.  For the solution~A,  the relative proper motion 
is much slower.  This means that using present instrumentation, one would have to wait 2.5 times longer, 
perhaps $\gtrsim 11$~years taking account of the fact that the host is somewhat fainter for solution A.  
However, in either case, the lens can be surely resolved from the source at first light on 30~m class 
telescopes of the next generation.  
Once the lens is resolved and its brightness is measured, the degeneracy can be checked using the 
additional constraint of the lens brightness, which is  $H\sim 22.0$ and $H\sim 20.1$ for the 
solutions A and B, respectively.  However, we note that the constraint of the lens brightness is 
relatively week because the masses and the magnitudes of the lens expected from the two solutions 
overlap in a wide range.  
Although future high-resolution followup observations may resolve 
the degeneracy between the solutions A and B, it will not be possible to resolve the close/wide 
degeneracy with any existing or proposed instrument.

\section{Summary and Conclusion}\label{sec:seven}

We carried out an analysis of KMT-2019-BLG-1339, for which a partially covered short-duration anomaly
appeared in the light curve.  Analysis indicated that the anomaly was generated a low-mass object 
accompanied to the lens.  However, accurate interpretation of the anomaly was prevented by two types 
of degeneracy, in which one originated from the ambiguity in $\rho$ and the other was the close/wide 
degeneracy.  The former degeneracy, finite-source degeneracy, resulted in ambiguities in both $s$ and $q$, 
and the latter degeneracy caused ambiguity only in $s$.  A Bayesian analysis using the Galactic priors 
yields that the masses of the lens components were $(M_1, M_2)\sim (0.27^{+0.36}_{-0.15}~M_\odot, 
11^{+16}_{-7}~M_{\rm J})$ and $\sim (0.48^{+0.40}_{-0.28}~M_\odot, 1.3^{+1.1}_{-0.7}~M_{\rm J})$ for the 
two sets of solutions, indicating that the lens comprises an M dwarf and an Jovian-mass planet or an 
object near the planet/brown dwarf boundary.  We estimated that the lens was located at distances of 
$\dl \sim 7.2^{+1.1}_{-1.3}$~kpc and $\sim 6.1^{+1.3}_{-1.6}$~kpc 
according to the individual solutions.  The finite-source degeneracy can be lifted from future observations 
using high-resolution instrument because the relative proper motions expected from the degenerate solutions 
are widely different.

\acknowledgments
Work by CH was supported by the grants  of National Research Foundation of Korea 
(2017R1A4A1015178 and 2019R1A2C2085965).
Work by AG was supported by the JPL grant 1500811.
The OGLE project has received funding from the National Science Centre, Poland, grant
MAESTRO 2014/14/A/ST9/00121 to AU.
This research has made use of the KMTNet system operated by the Korea
Astronomy and Space Science Institute (KASI) and the data were obtained at
three host sites of CTIO in Chile, SAAO in South Africa, and SSO in
Australia.

\end{document}